\documentclass[12pt,aps,prd,tightenlines,groupedaddress,preprint,floatfix,nofootinbib]{revtex4-1}

\usepackage{graphicx}
\usepackage{amssymb}
\usepackage{epstopdf}

\newcommand{\PRE}[1]{{#1}} 

\def\be{\begin{equation}}
\def\ee{\end{equation}}
\def\bea{\begin{eqnarray}}
\def\eea{\end{eqnarray}}

\def\tev{\, {\rm TeV}}
\def\gev{\, {\rm GeV}}

\newcommand{\gsim}{\lower.7ex\hbox{$\;\stackrel{\textstyle>}{\sim}\;$}}
\newcommand{\lsim}{\lower.7ex\hbox{$\;\stackrel{\textstyle<}{\sim}\;$}}

\newcommand{\pb}{\rm pb}

\newcommand{\Dsl}[1]{\slash\hskip -0.20 cm #1}

\begin{document}

\setlength{\baselineskip}{0.25in}

\setcounter{footnote}{0}
\setcounter{page}{1}
\setcounter{figure}{0}
\setcounter{table}{0}


\preprint{UH-511-1216-13, CETUP2013-015}

\title{\PRE{\vspace*{0.8in}}
WIMPy Leptogenesis With Absorptive Final State Interactions
\PRE{\vspace*{0.3in}}
}
\author{Jason Kumar}
\affiliation{Department of Physics and Astronomy, University of Hawaii, Honolulu, HI 96822 USA
\PRE{\vspace*{.5in}}
}

\author{Patrick Stengel}
\affiliation{Department of Physics and Astronomy, University of Hawaii, Honolulu, HI 96822 USA
\PRE{\vspace*{.5in}}
}

\begin{abstract}
\PRE{\vspace*{.3in}}
We consider a class of leptogenesis models in which the lepton asymmetry
arises from dark matter annihilation processes which violate $CP$ and
lepton number.  Importantly, a necessary one-loop contribution
to the annihilation matrix element arises from absorptive final state interactions.
We elucidate the relationship between this one-loop contribution and
the $CP$-violating phase.
As we show, the branching fraction for dark matter
annihilation to leptons may be small in these models, while still generating the necessary
asymmetry.
\end{abstract}


\maketitle

\section{Introduction}

The baryon asymmetry of the universe (BAU) is both evident in observation~\cite{Kolb:1990,Cline:2006} and well-motivated
theoretically at a variety of scales and epochs.
The existence of non-baryonic dark matter (DM) is also well established by observational signatures of its
gravitational interactions at several different scales~\cite{Kolb:1990,Bertone:2005,SDSS:2006,WMAP:2011}.
Calculations of primordial light element abundances
predicted by big bang nucleosynthesis (BBN)~\cite{Cyburt:2008}, recent observations of cosmic microwave background (CMB)
anisotropy by the Wilkinson Microwave Anisotropy Probe (WMAP)~\cite{WMAP:2011} and Planck satellite~\cite{Ade:2013zuv},
the galaxy power spectrum obtained by the
Sloan Digital Sky Survey (SDSS)~\cite{SDSS:2006}, and a variety of other data create a combined picture indicating that the
densities of baryonic and dark matter in our universe are
\bea
\Omega_b h^2 &\sim & 0.022  ,
\nonumber\\
\Omega_{DM} h^2 &\sim & 0.12 .
\eea
Both BAU and DM are strong motivations for physics beyond the Standard Model (SM), as neither significant baryon number (B)
violation nor appropriate
non-luminous gravitationally interacting fields exist within the SM. The most common models that explain DM  provide a
Weakly Interacting Massive Particle (WIMP) to account for the observed density~\cite{Bertone:2005}.
A typical WIMP, with weak scale mass and couplings, will depart from equilibrium in the early universe when
self-annihilation freezes out, yielding roughly the correct relic DM density.  It is natural to wonder if the annihilation
process which determines the dark matter density can also yield a baryon
asymmetry.

Many models have been proposed to explain BAU~\cite{Cline:2006}, but generally, in order for a process to produce the
observed baryon asymmetry, the Sakharov conditions~\cite{Sakharov:1967dj} for baryogenesis must be satisfied.
The Sakharov conditions require:
\begin{itemize}
\item{a violation of baryon number (or lepton number (L), if the asymmetry
is generated above the electroweak phase transition (EWPT)),}
\item{a violation of $C$ and $CP$,}
\item{a departure from thermal equilibrium.}
\end{itemize}

Although decaying dark matter with a weak scale mass has been suggested as a model for baryogenesis \cite{Kohri:2009}, a WIMP framework which
satisfies the first two conditions would automatically accommodate the third. This mechanism of
``WIMPy baryogenesis" was recently proposed by Cui, Randall and Shuve \cite{Cui:2012}. In a typical WIMPy model, dark matter, denoted $X$
with $m_X \sim~\tev$,
annihilates to an additional new weak scale field, denoted $H$, and a quark (or lepton) via $CP$-violating interactions.
The $CP$-violating phase arises from
interference between tree-level and loop diagrams. $H$ then subsequently decays into light particles, including particles which
are uncharged under Standard Model gauge symmetries, thus sequestering any negative baryon (or lepton) asymmetry.
Between the time $X$ begins to depart from equilibrium, $T \sim m_X$, and when
$X$ freezes out with the correct relic density, the correct baryon asymmetry can also be produced.
If DM annihilates to leptons, then any lepton asymmetry is transferred to baryons through electroweak sphalerons at temperatures above the EWPT.

While in Ref.~\cite{Cui:2012}, UV-complete models mediated by weak scale pseudoscalars are presented,
generalization to an effective field theory (EFT) was
more recently shown by Bernal, Josse-Michaux and Ubaldi~\cite{Bernal:2012}. The EFT WIMPy model achieves results similar to a UV-complete
model, with DM annihilating to $H$ and quarks through all possible dimension six operators. This allows one to
focus on aspects of the generation of a
baryon asymmetry which can arise from many different UV completions.

Subsequent work~\cite{Bernal:2013bga} provided a general analysis of how washout processes can erase an asymmetry.
The key constraint on WIMPy baryogenesis models, either constructed in a UV-complete or
EFT formalism, is that the couplings which must be added to generate needed one-loop contributions also introduce
dangerous tree-level washout processes.  These processes are ``dangerous" because they are not Boltzmann-suppressed,
even when some of the external particles are heavy.  As a result, features of the model must be chosen to render
these processes less dangerous.
This work focused on the particular case where $m_H \sim m_X$, and found a class of models in which
$H$ may decay entirely into Standard Model particles, while still generating the required baryon asymmetry.
For these processes, the asymmetry is largely generated after the washout processes are
frozen out.

In this work, we consider WIMPy leptogenesis\footnote{If dark matter
annihilation directly produces a baryon asymmetry, then
$H$ would have to be charged under $SU(3)_{qcd}$.  Models of this form are constrained by direct searches for color-charged particles
at colliders~\cite{ColorChargedSearches}, and Standard Model quarks produced by the decay
of $H$ would affect the baryon asymmetry.  Though such models are
certainly possible, we will for simplicity focus on the case where dark matter annihilation directly produces a lepton asymmetry,
along with a particle $H$ whose decay does not affect the baryon or lepton asymmetry. }~\cite{Fukugita:1986hr},
in which dark matter annihilation directly produces a lepton asymmetry which is converted into a
baryon asymmetry by electroweak sphalerons.  We focus on an effective field theory approach, in which dark matter annihilation can be parametrized
in terms of a set of effective operators.
In contrast to previous work, here we will obtain a $CP$-violating phase at tree-level from the interference between dimension six operators.
We will find that one-loop corrections
effects are necessary for generating an asymmetry, but they can be sequestered in absorptive terms in the final state interactions of $H$.
Because these terms do not directly
affect either the Standard Model or dark sector, we will be able to relax some of the parameter space constraints typically found in
WIMPy baryogenesis models.

We will find that some dangerous tree-level washout processes, which are often introduced by the interactions needed
to obtain $CP$-violation, are not present in the models we consider.  Moreover, the ratio of $CP$-violating to $CP$-conserving
interactions is not controlled by the scale of new physics.  As a result, one can generate the correct baryon asymmetry for
$\sim 1.5~\tev$ dark matter even if the scale of new physics is as large as $10~\tev$.

The outline of the paper is as follows.  In section II, we review the reason why a one-loop contribution
is necessary, and describe our class of models.  In section III we describe the solution of the Boltzmann
equations which govern the dark matter and lepton densities.  In section IV, we present our results and
compare to those of other models of WIMPy baryogenesis.  We conclude with a discussion of our results
in section V.

\section{Our Model}

We will consider models with a single dark matter candidate $X$ (stabilized by a $Z_2$ symmetry)
whose annihilation can be modeled with a
set of effective four-point operators.\footnote{In some models of~\cite{Bernal:2012} it is necessary
to use a $Z_4$ symmetry, since one of the final state particles is also charged under this stabilizing
symmetry.  That is not the case here, unless the effective operator respects $SU(2)_L$. In that case,
we would also need a $Z_4$ symmetry under which $X$ and $H$ are charged, in order to protect $H$ from
Standard Model decays which could wash out the
asymmetry.}
Each effective operator can be written as a product of an
initial state dark matter bilinear and a final state bilinear.  This final state bilinear couples
a Standard Model lepton ($L$) and an exotic fermion ($H$).
In order for the annihilation process to contribute to a lepton asymmetry,
the effective operators must violate $C$, $CP$ and $L$.

We can already constrain the set of effective operators which are relevant.  For dark
matter annihilation to contribute to a baryon asymmetry, there must be interference between two matrix
element terms with different phases (otherwise, any $CP$-violating phase would cancel in the squared
matrix element).  At lowest dimension, there are only two sets of fermionic
dark matter bilinears which can interfere in an annihilation process~\cite{Kumar:2013iva}:
\begin{itemize}
\item{$\imath \bar X \gamma^5 X$ and $\bar X \gamma^0 \gamma^5 X$ can both annihilate an $S=0$, $L=0$, $J=0$
($CP$-odd) initial state.}
\item{$\bar X \gamma^i X$ and $\bar X \sigma^{0i} X$ can both annihilate an $S=1$, $L=0$, $J=1$
($CP$-even) initial state.}
\end{itemize}
If the dark matter is spin-0, then there are no dimension 2 or dimension 3 bilinears which can interfere.
Moreover, if dark matter is a Majorana fermion, the second set of bilinears above vanish.

We will thus consider two sets of effective operators:
\bea
{\cal O}_{S=0} &=& {\lambda_1 \over 2M_*^2} (\imath \bar X \gamma^5 X) (\bar H P_L L)
+ {\lambda_1^* \over 2M_*^2} (\imath \bar X \gamma^5 X) (\bar L P_R H)
\nonumber\\
&\,&+ {\lambda_2 \over 2M_*^2} (\bar X \gamma_\mu \gamma^5 X) (\bar H \gamma^\mu P_L L)
+ {\lambda_2^* \over 2M_*^2} (\bar X \gamma_\mu \gamma^5 X) (\bar L \gamma^\mu P_L H )
\nonumber\\
{\cal O}_{S=1} &=& {\lambda_3 \over M_*^2} (\bar X \gamma^\mu X) (\bar H \gamma_\mu P_L L)
+ {\lambda_3^* \over M_*^2} (\bar X \gamma^\mu X) (\bar L \gamma_\mu P_L H)
\nonumber\\
&\,&+ {\lambda_4 \over M_*^2} (\bar X \sigma^{\mu \nu} X) (\bar H \sigma_{\mu \nu} P_L L)
+ {\lambda_4^* \over M_*^2} (\bar X \sigma^{\mu \nu} X) (\bar L \sigma_{\mu \nu} P_R H)
\label{eq:operator}
\eea
where $L$ is a Standard Model lepton and $H$ is an exotic field with no lepton number.  Note
that these operators cannot interfere with each other, as they annihilate initial states with
different spin and/or orbital angular momentum.  We may thus treat each operator separately.
Both operators are maximally $C$-violating.  Since both operators will yield similar results,
we focus on the case of ${\cal O}_{S=0}$.

The quantum numbers of $H$ depend on whether dark matter annihilates to a charged or neutral lepton
(we have assumed for simplicity that dark matter annihilates to a left-handed lepton; similar
results can be obtained if dark matter annihilates to a charged right-handed lepton).
The quantum numbers of the new fields for all cases are summarized in Table~\ref{tab:particles}.
We assume our effective operators do not respect $SU(2)_L$ and $U(1)_Y$, in order to ensure that lepton number
is not washed out by electroweak scale interactions between $H$ and Standard Model particles.\footnote{The authors are grateful to
B.~Garbrecht for pointing out these potentially large washout terms, discussed in
\cite{Garbrecht:2013}.  Note, these washout terms can also be forbidden if the discrete symmetry group
is enlarged to a $Z_4$ under which $H$ is charged.} Thus, our effective operator couplings must scale with the vev of the Higgs field.

\begin{table}[hear]
\centering
\begin{tabular}{|c|c|c|c|c|}
\hline
 Fields & $SU(2)_L$ &  $ Q_{U(1)_Y} $  & $ Q_{U(1)_L}$ & $\mathbb {Z}_2$ \\
\hline
$X$ & 1 & 0 & 0 & - \\
\hline
$P_L L = l_L$ & $\Box$ & -1/2 & +1 & +  \\
$H$ & $ 1 $ & 0 & 0 & + \\
\hline
$P_L L = \nu_L$ & $\Box$ & -1/2 & +1 & +  \\
$H$ & $ 1 $ & 0 & 0 & + \\
\hline
\end{tabular}
\caption{Particle Content}
\label{tab:particles}
\end{table}

The relative phase between the left-handed and right-handed components of
$X$, $L$ and $H$ can be fixed by requiring that they all have real mass eigenvalues.  The only phase rotations left
are non-chiral rotations of these fields, which can be used to absorb any overall phase
of the coefficients $\lambda_{1,...,4}$.
Note that the top line of eq.~\ref{eq:operator} is $CP$-invariant if $\lambda_1$ is purely
imaginary, while the last three lines are $CP$-invariant if $\lambda_{2,3,4}$ are purely real.  We thus
see that a $CP$-violating term must be proportional to $Re(\lambda_1 \lambda_2^*)$ or
$Im(\lambda_3 \lambda_4^*)$.

\subsection{One-loop corrections and absorptive interactions}

Consider the annihilation processes $XX \rightarrow Y$ and $XX \rightarrow \bar Y$, where
$Y$ represents any multi-particle final state, and $\bar Y$ is the $CP$-conjugate final state.
We may write the quantum matrix element for these processes as
\bea
{\cal M}_{XX \rightarrow Y} &=& {\cal M}_{XX \rightarrow Y}^{CP} + {\cal M}_{XX \rightarrow Y}^{CPV} ,
\nonumber\\
{\cal M}_{XX \rightarrow \bar Y} &=&\pm \left( {\cal M}_{XX \rightarrow Y}^{CP} - {\cal M}_{XX \rightarrow Y}^{CPV} \right) ,
\eea
where ${\cal M}^{CP}$ and ${\cal M}^{CPV}$ are the $CP$-invariant and $CP$-violating terms in the matrix element, respectively.
The sign of ${\cal M}_{XX \rightarrow \bar Y}$ is determined by the $CP$ transformation properties of the
initial state.
The final state asymmetry is then governed by the relation
\bea
\sigma_{XX \rightarrow Y} - \sigma_{XX \rightarrow \bar Y} &\propto & Re \left[{\cal M}_{XX \rightarrow Y}^{CP}
({\cal M}_{XX \rightarrow Y}^{CPV} )^* \right].
\eea
In order to generate a final state asymmetry, it is necessary that:
\begin{itemize}
\item{there exist both $CP$-invariant and $CP$-violating contributions to the matrix element.}
\item{the relative phase between the $CP$-invariant and $CP$-violating amplitudes differs
from $\pm \pi /2$.}
\end{itemize}
The first requirement above is satisfied by interference between two terms (parametrized by
coefficients $\lambda_1$ and $\lambda_2$) in the operator ${\cal O}_{S=0}$.

But for the $XX \rightarrow \bar H L$ matrix element generated by the operators in eq.~\ref{eq:operator},
the $CP$-invariant part is purely real and the $CP$-violating part is purely imaginary.
This is a result of the optical theorem, and as the second
point above indicates, implies that there will be no observable consequence to $CP$-violation.

The above line of reasoning leads to the usual result indicating that the generation
of an asymmetry from dark matter
annihilation requires interference between tree-level and one-loop diagrams.
The one-loop diagrams then generate a relative phase from regions of phase space where the
intermediate particles go on-shell, again as a result of the optical theorem.
The important point, however, is that the one-loop contribution
is {\it not} needed to provide a $CP$-violating matrix element; it is needed to generate the correct
phase between the $CP$-violating and $CP$-invariant terms in the matrix element.

But complex matrix element phases can also arise from final state absorptive interactions.
Within the effective operator approach, this one-loop correction is already present in
the $H$ external leg correction.  Assuming $H$ is unstable, its fully-corrected
propagator will have an imaginary contribution which is proportional to the total decay
width, $\Gamma_H$.  This imaginary contribution will be sufficient to generate a relative
phase between the $CP$-invariant and $CP$-violating amplitudes which differs from $\pm \pi /2$,
yielding a lepton asymmetry.

To be concrete, we will consider the case where $H$ is unstable
and decays through $H \rightarrow H' \phi$, where $H'$ is a fermion and $\phi$ is a
scalar (for simplicity we will assume that $m_{H'}, m_{\phi} \ll m_H$, and that the
Standard Model lepton $L$ is either stable, or has a much longer lifetime than $H$).
It is easy to see why treating $H$ as an unstable particle allows us to generate an
asymmetry between the total cross sections for the process $XX \rightarrow \phi^* \bar H' L$ and
$XX \rightarrow \phi \bar L  H'$.
Consider the operator ${\cal O}_{S=0}$ where we assume
$\lambda_{1,2}$ are real.  In this case, $\lambda_1$ is the coefficient of the $CP$-violating
operator which couples to the right-handed Weyl spinor $H_R$, while $\lambda_2$ is the
coefficient of the $CP$-invariant operator which couples to the left-handed Weyl spinor
$H_L$.  We will, for simplicity, assume that
$H$ can only decay from the left-handed helicity (as with Standard Model fermions), through
a $CP$-invariant operator
\bea
{\cal O}_H &=& |g| (\phi^* \bar H' P_L H + \phi \bar H P_R H').
\eea
If $\Gamma_H / m_H$ is sufficiently small, the dark matter annihilation amplitude will have an intermediate
$H$ which will be approximately on-shell.
We then
see that the $CP$-violating amplitude for $XX \rightarrow \bar H_R L_L \rightarrow \phi \bar H'_R  L_L$
depends on the helicity-flip term in the propagator of $H$, while the $CP$-invariant amplitude
for $XX \rightarrow \bar H_L L_L \rightarrow \phi \bar H'_R  L_L$ depends on the helicity-preserving
term in the $H$ propagator.  We can write the corrected $H$ propagator as~\cite{Kniehl:2008cj}
\bea
S(\Dsl p) = {{\Dsl p}_H + (m_H - \imath \Gamma_H /2) \over p_H^2 - m_H^2 -\imath m_H \Gamma_H },
\eea
where we see that the relative phase between the $CP$-violating and $CP$-invariant matrix elements
arises from the $-\imath \Gamma_H /2$ contribution to the helicity-flip term of the propagator.

The necessity of a one-loop contribution is
already familiar from previous work on WIMPy baryogenesis.
The difference in this work is that, unlike previous cases, here the one-loop correction is not the source of
$CP$-violation; $CP$-violation arises from the interference of two tree-level effective operators, and the
one-loop propagator correction only changes the relative phase between those terms in the matrix element.

This difference has important phenomenological consequences.
In models where the $CP$-violating phase is generated from loop diagrams, the required
additional field content and vertices typically introduce new tree-level washout processes which can
erase the asymmetry.  This typically results in a more constrained parameter space.  In our
example, however, since the one-loop contributions are sequestered from the Standard Model
and dark sectors, no new tree-level washout processes are introduced.  Moreover, the absorptive
terms can arise from strongly-coupled physics, even if the actual dark matter-Standard Model matter
interactions are perturbatively calculable.

\section{Calculation of the Baryon Asymmetry}

The tree-level cross section for dark matter annihilation is given by
\bea
\sigma_{tree}^{XX \rightarrow \bar H L \rightarrow \phi^*\bar H' L } v &=& { s\over 16\pi M_*^4}   \left\{ |\lambda_1|^2
+ { 4Im(\lambda_1^* \lambda_2) m_H m_X  \over s} \right.
\nonumber\\
&\,& \left.+ |\lambda_2|^2
\left[{4\over 3}\left(1 - {4m_X^2 \over s}\right) +{2m_H^2 \over 3s} +{4m_H^2 m_X^2 \over 3s^2}  \right] \right\}
\left[ 1 -{m_H^2 \over s} \right]^2 ,
\eea
where $\sqrt{s}$ is the energy in center-of-mass frame and for simplicity we assume $m_{H'}, m_\phi, m_L \ll m_X$.
At tree-level, the cross section for the conjugate
process $XX \rightarrow \bar L H \rightarrow \phi \bar L H' $ is the same.  But when one includes loop-corrections
to the $H$ propagator, one finds an asymmetry in the annihilation cross sections:
\bea
(\sigma^{XX \rightarrow \phi^* \bar H' L } - \sigma^{XX \rightarrow \phi \bar L H' } ) v &=&
\Gamma_H   {  Re( \lambda_1 \lambda_2^* ) m_X \over  4\pi M_*^4 }
\left[1- {m_H^2 \over s }\right]^2
\eea
where $\Gamma_H$ is the decay width of $H$, and we have assumed the narrow-width approximation.
If we define $\epsilon$ as the ratio of the cross section asymmetry to symmetric part:
\bea
\epsilon &\equiv& {\sigma^{XX \rightarrow \phi^* \bar H' L } - \sigma^{XX \rightarrow \phi \bar L H' }
\over \sigma^{XX \rightarrow \phi^* \bar H' L } + \sigma^{XX \rightarrow \phi \bar L H' } }
\eea
then we find $\epsilon \sim \Gamma_H / m_{H,X}$.
Assuming $m_X$ and $m_H$ are comparable, the narrow-width approximation would
be largely valid even for models with a cross section asymmetry as large as ${\cal O}(10\%)$.

The effective operator approximation will be largely valid if $m_X \ll M_*$.
To keep the heavy mediator effectively
decoupled from low-energy physics, we set $M_* = 10~\tev$.

\subsection{The Boltzmann equation}

We can write the Boltzmann equations in terms of dimensionless variables $ x = m_X / T$ and $Y = n / s $,
where $n$ is the number density and $s$ is the entropy density. Assuming an adiabatic process, the entropy $S$ should be constant,
and $Y$ is essentially a comoving number density.
We will assume that
$H'$ and $\phi$ are light particles which remain in equilibrium throughout the
relevant cosmological epoch, allowing us to make the approximation
$Y_\phi = Y_{\phi^*}= Y_{\phi_{eq.}}$, $Y_{H'} = Y_{\bar H'} = Y_{H'_{eq.}}$.
$L$ is a Standard Model lepton which is also light, but it will depart from
equilibrium due to the generated lepton asymmetry.  But this departure from
equilibrium will be small when the lepton asymmetry is small compared to the
total lepton density.  We define $Y_{\Delta L} \equiv Y_L - Y_{\bar L}$
as the asymmetry in $L$, which is either a charged lepton or neutrino of a single
generation, and we assume all generations have the same asymmetry.
We can assume $Y_L + Y_{\bar L} \simeq 2Y_{L_{eq}}$.
For dark matter annihilation to any Standard Model
fermion/antifermion pair, including the subdominant $CP$- and $L$-violating annihilations that are the source of
leptogenesis, the coupled Boltzmann equations are~\cite{Kolb:1990}:
\bea
{x^2 H(m_X) \over s(m_X) } {dY_X \over dx}  &=& - \langle \sigma_A v \rangle (Y_X^2 - Y_{X_{eq}}^2 ) ,
\label{eq:BoltzmannDM}
\\
{x^2 H(m_X) \over s(m_X) } {dY_{\Delta L}^{inj} \over dx}  &=&  {1 \over 2}\left[ \langle \sigma_{XX \rightarrow
\phi^* \bar H' L } v \rangle  \right]
(Y_X^2 - Y_{X_{eq}}^2 Y_L / Y_{L_{eq}} )
\nonumber\\
&\,& -{1 \over 2}\left[ \langle \sigma_{XX \rightarrow \phi \bar L H' } v \rangle   \right] (Y_X^2 - Y_{X_{eq}}^2 Y_{\bar L }/ Y_{\bar L_{eq}} )
\nonumber\\
&\,&
-\langle \sigma_{XL \rightarrow \phi X H' } v \rangle Y_X (Y_{L} - Y_{L_{eq}} )
+\langle \sigma_{X\bar L \rightarrow \phi^* X \bar H' } v \rangle Y_X (Y_{ \bar L} -Y_{\bar L_{eq}})
\nonumber\\
&\,& +...,
\label{eq:BoltzmannAsym}
\eea
where we have assumed the dark matter is a Majorana fermion and
the ``$+...$" terms involved suppressed $2 \rightarrow 3$ processes in which there is no
on-shell resonance.
$H(T)$ is the Hubble parameter  at temperature $T$ given a flat, radiation-dominated early universe.
The equilibrium rates for the relevant $3 \rightarrow 2$ processes are equal to the equilibrium rates
for the reverse $2 \rightarrow 3$ processes as a result of detailed balance.  The actual rates for
out-of-equilibrium $3 \rightarrow 2$ processes are determined by rescaling the equilibrium rates by the
ratio of the actual incoming particle densities to the equilibrium densities.  $dY_{\Delta L}^{inj} / dx$
is the rate at which a lepton asymmetry is injected by annihilation processes, not including the effects
of electroweak sphalerons.\footnote{Note, if the coupling is proportional to the Higgs vev, then
there can also be $2 \leftrightarrow 4$ processes in which a Higgs boson is produced. We disregard these
processes for simplicity, but they will not change the result significantly because they only rescale the
inclusive cross-section by a factor proportional to the phase space integration.  This rescaling can be
absorbed into the couplings. Alternatively, the charges of $H$ could be chosen such that the contacts
operators respect $SU(2)_L$ and $U(1)_Y$, in which case there is no Higgs coupling.
In this case, $H$ and $X$ could both be protected from contact
with the Standard Model with charges under a $Z_4$ discrete
symmetry~\cite{Bernal:2012}, but $X$ would have to be Dirac due to its imaginary charges. Our numerical results do not change
appreciably in this case.}

We can then rewrite the second equation Boltzmann equation as
\bea
{x^2 H(m_X) \over s(m_X) } {dY_{\Delta L}^{inj} \over dx}  &\sim& \langle \sigma_{XX}^{CPV} v \rangle (Y_X^2 - Y_{X_{eq}}^2)
- \langle \sigma_{XX}^{CP} v \rangle Y_{X_{eq}}^2 Y_{\Delta L} / Y_{L_{eq}}
- \langle \sigma_{XL}^{CP} v \rangle Y_X Y_{\Delta L} ,
\label{eq:SimplifiedBoltzmannAsym}
\eea
where
\bea
\langle \sigma_{XX}^{CP} v \rangle &\equiv & {1\over 2} \left[
\langle \sigma_{XX \rightarrow \phi^* \bar H' L } v \rangle + \langle \sigma_{XX \rightarrow \phi \bar L H' } v \rangle \right] ,
\nonumber\\
\langle \sigma_{XX}^{CPV} v \rangle &\equiv & {1\over 2} \left[
\langle \sigma_{XX \rightarrow \phi^* \bar H' L } v \rangle - \langle \sigma_{XX \rightarrow \phi \bar L H' } v \rangle \right] ,
\nonumber\\
\langle \sigma_{XL}^{CP} v \rangle &\equiv & {1 \over 2} \left[
\langle \sigma_{XL \rightarrow \phi X H' } v \rangle + \langle \sigma_{X\bar L \rightarrow \phi^* X \bar H' } v \rangle \right] .
\label{eq:SimplifiedCrossSection}
\eea
We can see that the first term on the right-hand side of eq.~\ref{eq:SimplifiedBoltzmannAsym} drives the asymmetry, while the
last two terms tend to wash it out.  Note that these washout terms are both Boltzmann-suppressed.

In this scenario, the only washout processes are those generated from the original four-point
effective operators via crossing symmetry.
Since the necessary loop
contribution arises only from the correction to the $H$ propagator, the introduction of the particles
which appear in the loop does not yield new tree-level washout processes.
As a result, there are no dangerous washout processes (in the sense of~\cite{Bernal:2013bga}) which are not Boltzmann suppressed.

Note that there is an asymmetry in the process $\phi \bar L H' \leftrightarrow \phi^* \bar H' L$ after one subtracts the
real intermediate state (RIS) one-loop diagrams in which an intermediate $XX$ two-particle state goes on-shell (this RIS contribution is
already accounted for in process $XX \leftrightarrow \phi \bar L H', \phi^* \bar H' L$).
The rates for the processes $\phi \bar L H' \leftrightarrow \phi^* \bar H' L$ can be related to the
rates for the processes $XX \leftrightarrow \phi \bar L H', \phi^* \bar H' L$ using the $CPT$-theorem,
which implies that the rates for the inclusive processes $\phi \bar L H' \rightarrow~anything$ and
$\phi^* \bar H' L \rightarrow~anything$ are identical.
We have implicitly included the rate asymmetries for these $3 \leftrightarrow 3$ processes in the Boltzmann equation, which
thus satisfies detailed balance.  We therefore generate no lepton asymmetry when the dark matter
is in equilibrium, in contrast to previous models of leptogenesis \cite{Garbrecht:2013}. Although we can safely ignore finite number
density corrections to our calculation, we could equivalently
use the CTP formalism \cite{Beneke:2010} to manifestly demonstrate the generation of the asymmetry, but this is beyond the scope of our paper.

We have not specified the high-energy Lagrangian which generates the effective operators described.  A
particular UV model which generates these effective operators at low-energy may also generate other
effective operators which contribute to washout processes, for example, operators of the form $(\bar H P_L L)^2$.
However, this is a model-dependent question; there will exist UV-completions (for example, models where the
mediating particle is exchanged in the $t$- or $u$-channel) in which such operators are not generated at
tree-level.  In keeping with our use of effective field theory, we will not assume the existence of any
additional effective operators beyond the ones we have introduced.

Although it is necessary to assume that $H$ is unstable in order to generate a lepton asymmetry,
we nevertheless were able to assume that the width of $H$ is relatively narrow.  This implies
that we should be able write equivalent Boltzmann equations in which we treat $H$ as a metastable
particle which is initially in thermal equilibrium.

\subsection{The effect of electroweak sphalerons}

The expression for $dY_{\Delta L}^{inj} / dx$ in the Boltzmann equation
provides the source term in the differential equation for the lepton number density,
which is coupled to the baryon number density through sphalerons. Given a number of generations $ N_G $ and
arbitrary lepton number sources $f_i$ for each generation $i$, we can write the evolution of the baryon and
lepton number densities \cite{Burnier:2006}
\bea
{dn_B \over dt} &=& - \gamma (t) \left[ n_B + \eta (t) \displaystyle\sum_{i=1}^{N_G} n_{L_{i}} \right] ,
\nonumber\\
{dn_{L_{i}} \over dt} &=& - {\gamma (t) \over N_{G}} \left[ n_B + \eta (t) \displaystyle\sum_{i=1}^{N_G} n_{L_{i}} \right]
+ { f_i (t)} .
\eea

The functions $\eta (T)$ and $\gamma (T)$ are defined in terms of the temperature $T$, the
temperature-dependent Higgs field expectation value $v_{min}$, and the Chern-Simons diffusion
rate $\Gamma_{diff} (T)$:
\bea
\eta (T) &=& { \chi  \left( T \right) \over {1 - \chi  \left(  T \right)}} ,
\nonumber\\
\gamma (T) &=& N_{G}^{2} \, \rho \left(  T \right) \left[ 1 - \chi  \left(  T \right) \right]
 {\Gamma_{diff} (T) \over T^3} ,
\nonumber\\
\rho (T) &=& {3 \left[ 65 + 136 N_G + 44  N_G^2  + (117 + 72 N_G) \left( v_{min} \over T \right)^2 \right]
\over { 2 N_G \left[ 30 + 62 N_G + 20 N_G^2 + (54 + 33 N_G) \left( v_{min} \over T \right)^2 \right]}}
\nonumber\\
\chi (T) &=&  {4 \left[ 5 + 12 N_G + 4  N_G^2  + (9 + 6 N_G) \left( v_{min} \over T \right)^2 \right]
\over {  65 + 136 N_G + 44 N_G^2 + (117 + 72 N_G) \left( v_{min} \over T \right)^2 }}.
\eea

\begin{figure}[hear]
\center\includegraphics[width=\textwidth]{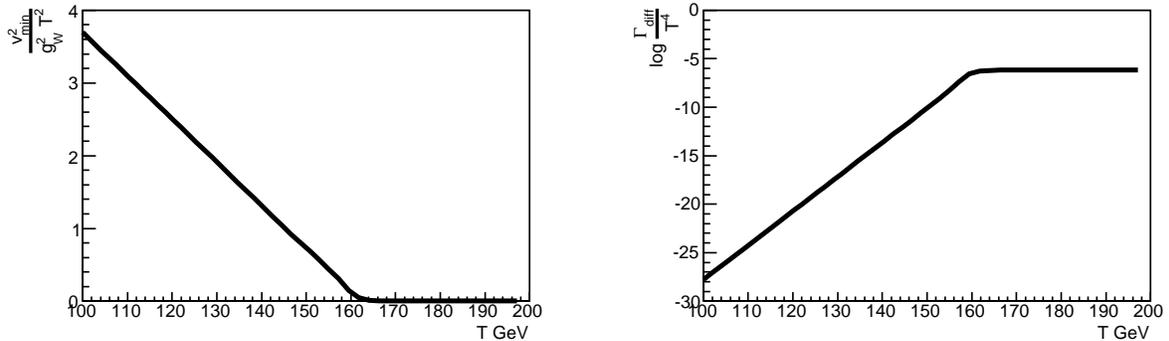}
\caption{Lattice calculation of $v_{min}^2 / (T^2 g_{weak}^2)$ (left) and $\log (\Gamma_{diff} /T^4)$
(right) through the transition region \cite{D'Onofrio:2012}.  For $T<140~\gev$, $\log (\Gamma_{diff} /T^4)$ is
extrapolated from analytical calculations deep in the broken phase \cite{Burnier:2006}, with a constant
rescaling to provide consistency with the lattice calculation for $T=140-155~\gev$.}
\label{fig:SphaleronStuff}
\end{figure}

The temperature dependence of both $v_{min}$ and $\Gamma_{diff}$ has been calculated on the lattice through the
electroweak transition region \cite{D'Onofrio:2012} and analytically deep in the broken phase \cite{Burnier:2006}.
The lattice and analytical calculations of $v_{min}$ are
consistent.
The lattice and analytical calculation of $\Gamma_{diff}$ differ by an order of magnitude in the range of overlap
($T=140-155~\gev$), but exhibit the same logarithmic slope.    We will use the  $\Gamma_{diff} (T)$
determined from lattice calculations for $T \geq 140~\gev$ and use the
analytical result for $T \leq 140~\gev$ after rescaling the analytical result by a constant factor
to provide consistency with lattice calculations in the region of overlap.  The $v_{min} (T)$ and $\Gamma_{diff} (T)$
which we use are plotted in figure~\ref{fig:SphaleronStuff}.

Now we recast the coupled equations for the comoving baryon and lepton numbers in terms of our dimensionless
source variables, noting $ N_G = 3 $
and assuming all lepton generations evolve equivalently.

\bea
x H(T) {dY_{\Delta B} \over dx} &=& - \gamma (T) \left[ Y_{\Delta B} + 3 \eta (T) Y_{\Delta L} \right]
\nonumber\\
x H(T) {dY_{\Delta L} \over dx} &=&  - {1 \over 3} \gamma (T) \left[ Y_{\Delta B} + \eta (T) Y_{\Delta L} \right]
+ x H(T) { dY_{\Delta L}^{inj} \over dx}
\eea

\section{Results}

We will assume that $m_H \leq 2m_X$, so that the process $\bar X X \rightarrow \bar H L$ is kinematically
allowed.  As in~\cite{Cui:2012,Bernal:2012,Bernal:2013bga}, we will assume $m_H \sim m_X$.
We assume that the reheating temperature of the universe is large enough that the dark matter
was in relativistic thermal equilibrium in the early universe ($x < 1$).
We then numerically solve the coupled Boltzmann/sphaleron rate equations, using equilibrium at $x=1$ as a boundary condition.

Sphaleron processes will start to decouple for temperatures $T \lesssim {\cal O}(100)~\gev$.
We will thus find that, for $m_X \gg 100~\gev$, the lepton asymmetry is generated when sphalerons
are active, and the magnitude of the baryon asymmetry is roughly the same as that of the lepton
asymmetry at late times.

For all of the models we consider, we take $\langle \sigma_A v \rangle = 1~\pb \gg
\langle \sigma_{\bar X X \rightarrow \bar H L} v \rangle$.  As a result, $X$ will annihilate rapidly
enough to ensure that its density does not exceed observational bounds, while the annihilation process
$\bar X X \rightarrow \bar H L, \bar L H$ will not significantly affect the dark matter density (though
it will impact the baryon asymmetry).  For the case where only the operator ${\cal O}_{S=0}$ is present,
the relevant parameters of the model are $m_X$, $m_H / m_X$, $\Gamma_H / m_H$ and $Re (\lambda_1 \lambda_2^*)$ (the last parameter
is replaced by $Im (\lambda_3 \lambda_4^*)$ in the case where only ${\cal O}_{S=1}$ is present).

\begin{figure}[hear]
\center
\includegraphics[width=\textwidth]{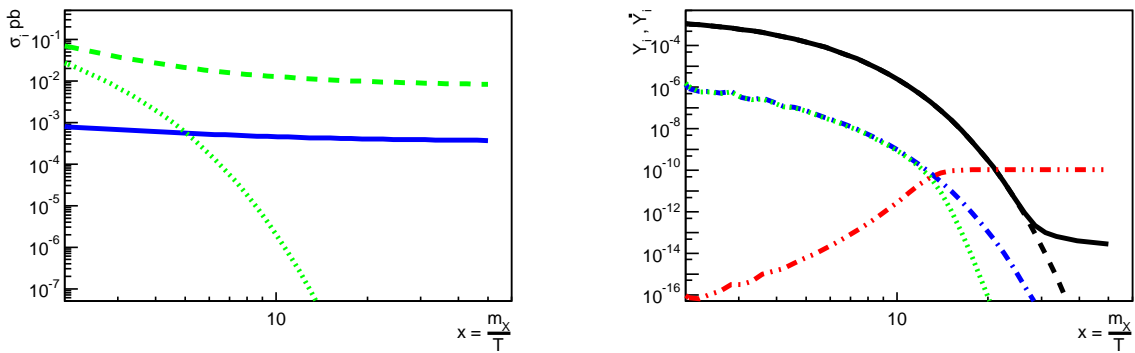}
\caption{The left panel shows the thermally-averaged cross sections $\langle \sigma_{XX}^{CP} v  \rangle$ (green dashed),
$\langle \sigma_{XX }^{CPV} v  \rangle$ (blue solid) and $\langle \sigma_{XL }^{CP} v  \rangle$ (green dotted) (as defined
in eq.~\ref{eq:SimplifiedCrossSection}).
The right panel shows the corresponding contrinbutions to the lepton source rate (sum of washout terms
in green dotted and source term in blue dash single dotted) as well as $Y_B$ (red dash double dotted), $Y_X$ (black solid) and $Y_{X_{eq}}$ (black dashed).
We have chosen parameters
$ m_{X} = 5~\tev$, $ m_{H} = 7~\tev$,  $\lambda_1 = \lambda_2 = 0.5 $, $\langle \sigma_A v \rangle = 1~\pb$,
$\Gamma_H / m_H = 0.1 $ . }
\label{fig:injection}
\end{figure}

We can then illustrate our results with some benchmark points.
In figure~\ref{fig:injection}, we plot the thermally-averaged cross sections for the processes
$XX \rightarrow \phi^* \bar H' L$ (both $CP$-invariant and $CP$-violating terms) and
$XL \rightarrow \phi H' X$ (the $CP$-invariant term) as a function of $x = m_X / T$.
We also plot the contribution of these terms
to the lepton source injection rate, as well as $Y_B$, $Y_X$ and $Y_{X_{eq}}$.
We have chosen the ``high-mass" benchmark parameters
$m_{X} = 5~\tev$, $m_{H} = 7~\tev$, $\lambda_1 =  \lambda_2 = 0.5  $,
$\langle \sigma_A v \rangle = 1~\pb$, {\bf $\Gamma_H / m_H = 0.1$}.
As we expected, the process $XL \rightarrow \phi H' X$ is kinematically-suppressed at low temperature.
Note, however, that we have assumed the narrow-width approximation; this approximation will break down
at sufficiently low temperatures, when the Boltzmann suppression required for the production of an
on-shell $H$ is larger than the cross section suppression when $H$ is off-shell.
But if we instead take $H'$ and/or $\phi$ to be massive, then even this off-shell process can
be suppressed, while on-shell processes will be unaffected.

\begin{figure}[hear]
\center\includegraphics[width=\textwidth]{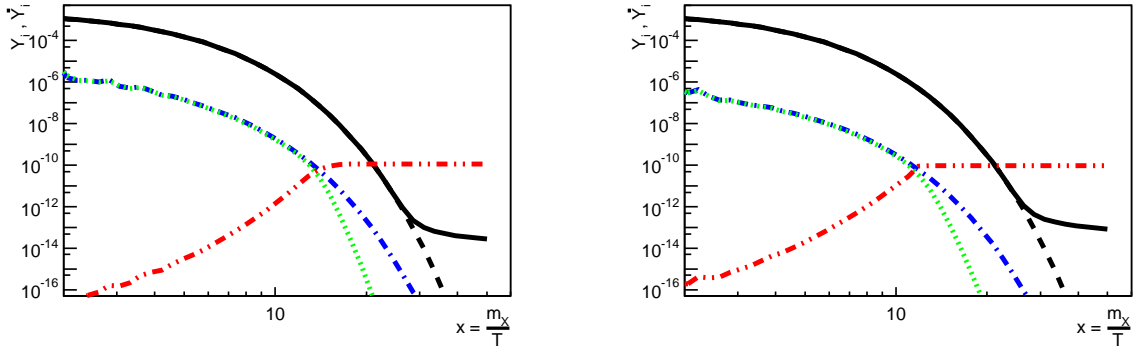}
\caption{ Rate contributions and Boltzmann equation solutions for our narrower-width (left panel) and low-mass (right panel)
benchmarks (colors same as in right panel of figure~\ref{fig:injection}). Parameters are summarized in table~\ref{tab:benchmarks}.
}
\label{fig:density}
\end{figure}

In figure~\ref{fig:density} we plot the $Y_B$, $Y_X$, $Y_{X_{eq}}$ and the lepton injection
rates (both source and washout terms)
as a function of $x = m_X / T$ for a narrower-width benchmark model with $m_{X} = 5~\tev$,
$m_{H} = 7~\tev$, $\lambda_1 =  \lambda_2 = 0.5 $, $\langle \sigma_A v \rangle = 1~\pb$, {\bf $\Gamma_H / m_H = 0.05$} (left panel)
and for a low-mass benchmark model with $ m_{X} = 1.5~\tev$,  $m_{H} = 2.2~\tev$, $\lambda_1 =  \lambda_2 = 1$,
$\langle \sigma_A v \rangle = 1~\pb$, {\bf $\Gamma_H / m_H = 0.1$} (right panel).
For all benchmark models, the couplings are chosen so that the final baryon asymmetry matches observation.
Note that the narrower-width benchmark is nearly identical to the high mass benchmark, only washout processes freeze out slightly later.
For the low-mass benchmark, sphalerons begin to decouple around when washout processes freeze out,
thus forcing a sharper freeze out of baryon number.
The parameters of these benchmark models are summarized in Table~\ref{tab:benchmarks}.

\begin{table}[hear]
\centering
\begin{tabular}{|c|c|c|c|c|c|c|}
\hline
benchmark & $m_X$ & $m_H$ &  $ \Gamma_H / m_H $  & $ \lambda_1 = \lambda_2 $ & $ \epsilon $ &
$ \langle \sigma_{XX \rightarrow \phi^* \bar H' L} v  \rangle  /  \langle \sigma_A v  \rangle $  \\
\hline
low-mass & $ 1.5~\tev $ & $ 2.2~\tev $ & $ 0.10 $ & $ 1.0 $ & $ 0.045 $  & $ 0.002 $ \\
\hline
high-mass & $ 5.0~\tev $ & $ 7.0~\tev $ & $ 0.10 $ & $ 0.5 $ & $ 0.045 $  & $ 0.008 $ \\
\hline
narrower-width & $ 5.0~\tev $ & $ 7.0~\tev $ & $ 0.05 $ & $ 1.0 $ & $ 0.022 $  & $ 0.033 $ \\
\hline
\end{tabular}
\caption{Benchmarks}
\label{tab:benchmarks}
\end{table}

It is interesting to note that the $CP$-violating part of the process $XX \leftrightarrow \bar L H , \bar H L$
begins to drive the
asymmetry near $x \sim 1$, i.e.~as soon as the dark matter becomes non-relativistic.  This may seem
counterintuitive, since the Sakharov conditions require a departure from thermal equilibrium and dark matter
freeze out occurs much later, near $x \sim 20-30$.  The point is that the dark matter density departs
slightly from the equilibrium density as soon as dark matter becomes non-relativistic.  At freeze-out,
dark matter stops tracking the equilibrium density and the departure from equilibrium becomes large.
One can see this simply by considering the Boltzmann equation (eq.~\ref{eq:BoltzmannDM}); if the dark matter density is equal
to the equilibrium density, then $dY / dx =0$.  This relation is satisfied if dark matter is relativistic
($Y \sim const.$), but is violated when dark matter is non-relativistic ($Y \sim x^{-3/2}e^{-x}$).  A slight
departure from equilibrium is necessary to provide the excess annihilation which drives $Y$ to smaller
values.  Although the departure from equilibrium is very small before freeze out, the dark matter density
at $x \sim 1$ is so much larger than at $x \sim 20$ that the driving contribution to the lepton
asymmetry is largest at small $x$.  However, at small $x$ the washout processes are also at their strongest and
the net asymmetry is quite small.  A large asymmetry begins to be generated as soon as the washout processes
begin to freeze out, which for these models typically happens near $x \sim 10$, as in~\cite{Bernal:2013bga}.

There are a few features which
distinguish our results from those of other WIMPy baryogenesis models.
First of all, in previous works, the interactions necessary to produce the needed one-loop diagrams also
introduce tree-level washout diagrams in which $X$ does not appear as an initial or final state.  These
washout processes are essentially processes of the form $\bar H L \leftrightarrow \bar L H$, and
can be significant even when $T < m_X$.  On the other hand, in the class of models
we consider, all washout diagrams have $X$ as a final state.

Note also that $\epsilon \sim \Gamma_H / 2m_{H,X}$, and
is independent of the mediator scale $M_*$.
This is in contrast with other WIMPy baryogenesis models, where
one typically finds $\epsilon \propto m_X^2 / M_*^2$~\cite{Bernal:2012}.
This difference has some interesting effects.  If we assume that $X$ constitutes all of the dark
matter, then the total cross section for the annihilation process $XX \rightarrow \bar H L, \bar L H$
is bounded by $\sim 1~\pb$.  In order to generate a large enough asymmetry, $\epsilon$ cannot be too
small; it appears that one would need $\epsilon \sim {\cal O}(0.01-0.1)$.  If $\epsilon \sim  (\lambda^2 /4\pi ) m_X^2 / M_*^2$
and $\lambda \sim 1$, this would require that $M_* \lsim 3m_X$.  By contrast, in our low-mass benchmark model, the
new physics scale $M_* $ can be much larger, since $\epsilon$ is independent of the scale of the new physics
in the effective operator.
As a result, the total $XX \rightarrow \bar H L, \bar L H$ cross section
can be much smaller than the $\langle \sigma_A v \rangle \sim 1~\pb$.

In all of the models we have considered, we have chosen $m_H \sim m_X$.  It is difficult
to find a successful model if one instead chooses $m_H \ll m_X$.  The reason is because
we have assumed the narrow-width approximation, in which $H$ acts as a resonance, and thus
require $\Gamma_H \ll m_H$. In such a model, since we would also still require $\epsilon \propto \Gamma_H / m_X$, a light $H$ field
would imply a small cross section asymmetry.

\section{Conclusions}

We have considered a model of WIMPy leptogenesis, in which a lepton asymmetry is generated by
dark matter annihilation processes which violate $C$, $CP$ and lepton number.  We have studied
in detail the necessity for one-loop contributions to the annihilation process.  In particular,
we have found the $CP$-violating terms may all arise at tree-level, while one-loop diagrams
may arise only in absorptive final state interactions in a sterile sector.

The advantage of this type of model is that it allows one to sequester the one-loop suppression from
the generation of a $CP$-violating phase.  In WIMPy baryogenesis models where this sequestration does
not occur, the introduction of one-loop terms implies the presence of new tree-level process which
are not Boltzmann suppressed and which can washout the baryon asymmetry.  In the class of models we
have considered, these dangerous washout processes do not occur.  Moreover, because the ratio of
$CP$-violating to $CP$-conserving terms in the cross section is independent of the scale of new
physics, we find that WIMPy models with $m_X \sim 1.5~\tev$ can work with a new physics scale
$M_*$ as large as $10~\tev$, and where the annihilation cross sections relevant for WIMPy leptogenesis are
much smaller than $1~\pb$.

It is interesting to consider prospects for probing these dark matter interactions experimentally.
Indirect detection may be feasible for WIMPy leptogenesis models in general,
but given our specific $CP$-violating processes, primary dark matter annihilation channels would likely
dominate over any measurable imprint our subdominant channel would leave on the cosmic ray spectrum.
Better prospects may lie with high energy/luminosity $e^+ / e^-$ colliders, which may be able to probe
lepton number- or lepton flavor-violating processes to which these operators could contribute.

For simplicity, we have focused on the approximation where, despite the presence of final state
absorptive interactions, the sterile particle $H$ can be treated as a narrow resonance produced
on-shell.  But this assumption is not required, and if the narrow-width approximation does not hold,
then all of the asymmetry-generating and washout processes would have to be fully treated as
$2 \rightarrow 3$ processes.  In this case, one would expect that one-loop suppression required for
$CP$-violating annihilation rates would be significantly reduced.  It would be interesting to study
this scenario concretely.

{\bf Acknowledgements}

We are grateful to S.~Pakvasa, X.~Tata, B.~Thomas and L.~Ubaldi for useful discussions.
This work is supported in part by Department of Energy grants DE-FG02-04ER41291 and DE-FG02-13ER41913.
We thank the organizers of TASI 2013 and the Center for Theoretical
Underground Physics and Related Areas (CETUP) in South Dakota
for their support and hospitality while this work was being completed.

\end{document}